\documentclass[aps,prb,twocolumn,groupedaddress,showpacs]{revtex4}
\usepackage{latexsym} 
\usepackage[dvips]{graphicx} 
\usepackage{amssymb}
\usepackage{amsmath}

\begin{document}

\title{Conductance Peak Height Correlations for a Coulomb-Blockaded Quantum Dot in a Weak Magnetic Field}

\author{Stephan Braig, Shaffique Adam, and Piet W. Brouwer}
\affiliation{Laboratory of Atomic and Solid State Physics, Cornell University, Ithaca, New York 14853-2501}
\date{January 21, 2003} 

\begin{abstract}
We consider statistical correlations between the heights of conductance peaks corresponding to two different levels in a Coulomb-blockaded quantum dot. Correlations exist for two peaks at the same magnetic field if the field does not fully break time-reversal symmetry as well as for peaks at different values of a magnetic field that fully breaks time-reversal symmetry. Our results are also relevant to Coulomb-blockade conductance peak height statistics in the presence of weak spin-orbit coupling in a chaotic quantum dot.
\end{abstract}
\maketitle

\section{Introduction}
Measurement of conductance peak heights in a Coulomb-blockaded quantum
dot is one of few experimental tools to access properties of
single-electron wavefunctions in quantum dots. Experimentally, 
the probability
distribution of the conductance peak heights in quantum dots with an
irregular shape was found to be in good agreement with predictions
from random-matrix theory (RMT),~\cite{kn:jalabert1992} both without magnetic field and in the
presence of a time-reversal symmetry breaking magnetic field.~\cite{kn:marcus,kn:chang} According to
random-matrix theory, wavefunctions
in a chaotic quantum dot have a universal distribution,
independent of details of the dot's shape or mean free
path,
and wavefunction elements are independently
gaussian distributed real or complex numbers 
depending on the presence or absence of time-reversal symmetry,
respectively.
There are no long-range
correlations within a chaotic wavefunction and no correlations between
different wavefunctions.~\cite{kn:berry,kn:mirlin}

It is known that non-universal
correlations between different wavefunctions and,
hence, correlations between conductance peak heights exist
in both ballistic and diffusive dots.~\cite{kn:foot1} In 
ballistic quantum dots, such correlations are the result of 
wavefunction scarring,~\cite{kn:heller} 
which causes a slow modulation of the variance $\langle
|\psi_{\mu}(\vec r)|^2 \rangle$ as a function of the level
index $\mu$ and the position $\vec r$, although the RMT prediction
for peak-height statistics remains valid for peaks at
nearby energies.~\cite{kn:narimanov} 
The scarring effect disappears in the limit of large
quantum dots, and is absent in quantum dots with scatterers smaller
than the Fermi wavelength. 
In disordered quantum dots (mean free path $l$ much smaller than
dot size $L$), correlations between conductance peak heights are found to be of relative order $({\Delta}/{E_{\rm T}}) \ln (L/l)$,~\cite{kn:mirlin,kn:blantermirlin} where $E_{\rm T}$ is the Thouless energy of the quantum dot and $\Delta$ the mean level spacing.

Here, we address two other mechanisms for peak-height correlations. On the one hand, we investigate correlations at a weak perpendicular magnetic field that only partially breaks time-reversal symmetry. These correlations follow from underlying correlations of wavefunctions, which were reported previously by Waintal, Sethna, and two of the authors.~\cite{kn:absw} On the other hand, we also find 
correlations between peaks corresponding to different wavefunctions at two different values of a large magnetic field that fully breaks time-reversal symmetry.
Unlike the two other causes for peak-height correlations, the source
of correlations under investigation in this paper is universal and survives in
the limit of large quantum dots. Furthermore, these correlations are
not only of direct experimental relevance when Coulomb-blockade peaks
are measured as a function of an external magnetic field, but our
results also pertain to the case of quantum dots with weak spin-orbit scattering. We elaborate on this aspect at the end of the paper.

Our paper is organized as follows: in Section \ref{sec:model}, we introduce the Pandey-Mehta Hamiltonian, which is the RMT Hamiltonian appropriate for our calculations. We then proceed in Section \ref{sec:corr} to formulate the problem in terms of orthogonal invariants of the Pandey-Mehta Hamiltonian and derive a general expression for the wavefunction correlator distribution function. This result is employed to calculate the actual peak height correlator distribution function whose first moment is compared to numerical RMT simulations, both for the case of a weak magnetic field (Sec. \ref{sec:wmf}) and different large magnetic fields (Sec. \ref{sec:lmf}). Finally, we apply our results to correlations in presence of spin-orbit coupling and to ``spectral scrambling'' in Section \ref{sec:appl}.

\section{RMT Model}
\label{sec:model}
\begin{figure}
\scalebox{0.6}{\includegraphics{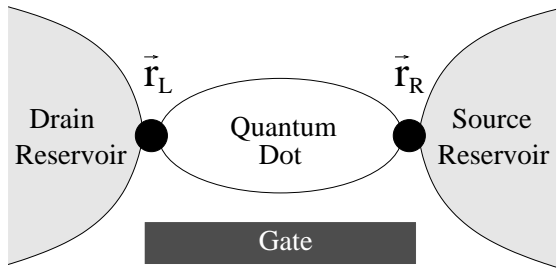}}
\caption{ \label{schematic} The quantum dot is connected to source and drain reservoirs by tunneling contacts at $\vec{r}_R$ and $\vec{r}_L$ and is capacitively coupled to a gate.}
\end{figure}

At temperatures $k_B T \ll \Delta$, the maximum conductance $G_{\mu}^{\rm peak}$ of a Coulomb-blockade conductance peak is a function of the wavefunction $\psi_{\mu}(\vec r)$ of the resonant state $|\mu \rangle$ only,~\cite{kn:beenakker,kn:albrgl}
\begin{equation}
  \label{eq:gpeak}
  G_{\mu}^{\rm peak}=
  \left(\frac{e^2}{h}
  \frac{V \Delta}{\kappa k_B T}\right)
  \frac{T_L |\psi_{\mu}(\vec r_L)|^2 T_R|\psi_{\mu}(\vec r_R)|^2}
  {T_L |\psi_{\mu}(\vec r_L)|^2 + T_R |\psi_{\mu}(\vec r_R)|^2}
  .
\end{equation}
Here, $V$ is the area of the quantum dot, $\kappa= \frac{3}{2} + \sqrt{2}$,
$\vec r_R$ and $\vec r_L$ are the positions of the tunneling
contacts connecting the dot to source and drain reservoirs (cmp. Fig. \ref{schematic}), and
$T_L$, $T_R \ll 1$ are the transmission probabilities of the contacts.
Equation (\ref{eq:gpeak}) is valid in the experimentally relevant
range of thermally
broadened conductance peaks,  $(T_{L} + T_{R}) \Delta 
\ll k_B T \ll \Delta$.
The index $\mu$ is defined with respect to
the orbital state. In the presence of spin degeneracy, each orbital
state gives rise to two conductance peaks, although
these two peaks do not need to appear in
succession.~\cite{kn:boh,kn:bug}
We express our results in terms of the distribution of the
dimensionless peak height $g_{\mu}$, 
\begin{equation}
  G_{\mu}^{\rm peak} = g_{\mu} \left(\frac{e^2}{h} \frac{\Delta}{\kappa k_B T}\right)
  \frac{T_L T_R}{T_L+T_R},
\end{equation}
and calculate the connected
part of the joint conductance peak height distribution for
two different levels $\mu$ and $\nu$,
\begin{equation}
  P_c(g_{\mu},g_{\nu}) = P(g_{\mu},g_{\nu}) - P(g_{\mu}) P(g_{\nu}).
\end{equation}
The single-peak distribution $P(g_{\mu})$ for the case of weak
magnetic fields was calculated by Alhassid {\em et
  al.}.~\cite{kn:alhassid} 

Within random-matrix theory,
the effect of a magnetic field is described by the Pandey-Mehta 
Hamiltonian~\cite{kn:pandeymehta}
\begin{equation}
  H(\alpha) = S + i \frac{\alpha}{\sqrt{2 N}} A,
  \label{pandey}
\end{equation}
where $S$ and $A$ are symmetric and antisymmetric $N\times  N$
matrices, respectively, with identical and independent
Gaussian distributions. The parameter $\alpha$ is proportional
to the magnetic field $B$,
\begin{equation}
  \alpha = \gamma \frac{e B V}{hc} \sqrt{\frac{E_T}{ \Delta}},
\end{equation}
where $\gamma$ is a constant of order unity that depends on the precise
geometry of the dot, for example $\gamma=\sqrt{\pi/2}$ for a diffusive disk of radius $L$ ($E_T=\hbar v_F l/L^2$) and $\gamma=\pi/\sqrt{8}$ for a ballistic disk with diffusive boundary scattering ($E_T=\hbar v_F/L$).~\cite{kn:frahm,kn:beenakkerRMT,kn:MishaPiet2002}

\section{Wavefunction Correlations and Peak Height Correlator}
\label{sec:corr}
The joint distribution of eigenvectors of the Pandey-Mehta Hamiltonian
(\ref{pandey}) is determined by the orthogonal
invariants $\rho_{\mu\nu}\equiv v_\mu^Tv_\nu$, where the superscript 
$T$ denotes transposition.~\cite{kn:absw} 
At fixed $\rho_{\mu\nu}$ and for large $N$, eigenvector 
components are distributed according to a multivariate Gaussian 
distribution with covariance matrix determined by the 
pair correlators
\begin{eqnarray}
  \label{variancesGOE}
  \langle v_{\mu,m}^* v_{\nu,n}\rangle_\rho = 
  \frac{\delta_{mn}\delta_{\mu\nu}}{N}, \;\;
  \langle v_{\mu,m} v_{\nu,n}\rangle_\rho = 
  \frac{\delta_{mn}\rho_{\mu\nu}}{N}.
\end{eqnarray}
The distribution of the orthogonal invariants is known for the
limiting cases $|\varepsilon_{\mu} - \varepsilon_{\nu}| \gg
\Delta$ or $\alpha \gg 1$, when their distribution is Gaussian
with zero mean and with variance~\cite{kn:absw}
\begin{eqnarray}
\label{rhovar} \langle |\rho_{\mu\nu}|^2\rangle &=& 
\frac{2\alpha^2(1+\delta_{\mu\nu})}{4\alpha^4+
\pi^2(\varepsilon_\mu-\varepsilon_\nu)^2/\Delta^2}.
\end{eqnarray}
Furthermore, if $|\varepsilon_{\mu} - \varepsilon_{\nu}| \gg \Delta$ or $\alpha\gg 1$, $|\rho_{\mu\mu}|^2$ and $|\rho_{\nu\nu}|^2$ are statistically
independent. Eq.~(\ref{rhovar}) is also valid in the limit $\alpha 
\ll 1$, if an additional average over the energy levels
$\varepsilon_{\mu}-\varepsilon_{\nu}$ is taken.
No analytical results are known for
the distribution of the orthogonal invariants
$\rho_{\mu\nu}$ when $\mu \neq \nu$, $\alpha$ is of order unity, and 
$|\varepsilon_{\mu} - \varepsilon_{\nu}| \lesssim \Delta$. 

Using the correspondence between the eigenvectors of the Pandey-Mehta
Hamiltonian and the wavefunctions in the quantum dot, we identify
$\psi_{\mu}(\vec r_L)$ with $v_{\mu,1}$ and $\psi_{\mu}(\vec r_R)$
with $v_{\mu,2}$. We are interested in the joint distribution of the
wavefunctions corresponding to the levels $\mu$ and $\nu$ and
abbreviate $x_1 = N |v_{\mu,1}|^2$, $x_2 = N |v_{\mu,2}|^2$, $y_1
= N |v_{\nu,1}|^2$, and $y_2 = N |v_{\nu,2}|^2$. To leading order
in $\rho_{\mu\nu}$, the connected part of an average of the
form $\langle x_1^k x_2^l y_1^m y_2^n\rangle_c
=\langle x_1^k x_2^l y_1^m y_2^n\rangle - \langle x_1^k x_2^l 
\rangle \langle y_1^m y_2^n\rangle$
 can be calculated
with the help of Wick's theorem and Eq.~(\ref{variancesGOE}),
\begin{eqnarray}
\left\langle x_1^k x_2^l y_1^m y_2^n\right\rangle_c &=&
  \langle|\rho_{\mu\nu}|^2 \rangle
  \left[ k^2 m^2\left\langle x_{1}^{k-1} x_{2}^{l}\right\rangle
  \left\langle y_{1}^{m-1} y_{2}^{n}\right\rangle 
  \right. \nonumber\\ &&
  \left. \mbox{} + l^2 n^2\left\langle x_{1}^{k}
  x_{2}^{l-1}\right\rangle
  \left\langle y_{1}^{m} y_{2}^{n-1}\right\rangle \right]
  \nonumber. \!\!
\end{eqnarray}
In the regimes $|\epsilon_\mu-\epsilon_\nu|\gg\Delta$ or $\alpha\gg 1$, this relation allows us to express the connected part of the joint
distribution function $P_c(x_1,x_2;y_1,y_2)
= P(x_1,x_2;y_1,y_2) - P(x_1,x_2) P(y_1,y_2)$ in terms of the
distribution functions $P(x_1,x_2)$ and $P(y_1,y_2)$ for elements
of a single eigenvector,
\begin{eqnarray}
  P_c(x_1,x_2;y_1,y_2) &=& \frac{2\alpha^2}{4\alpha^4+
\pi^2(\varepsilon_\mu-\varepsilon_\nu)^2/\Delta^2}
  \nonumber \\ &&\!\!\!\!\!\!\!\!\!\!\!\!\!\!\! \mbox{} \times
  \sum_{j=1}^{2} D_{x_j} P(x_1,x_2) D_{y_j} P(y_1,y_2),
  \label{Pc}
\end{eqnarray}
where $D_x\equiv\partial_x x\partial_x$. The single-wavefunction
distribution $P(x_1,x_2)$ for the Pandey-Mehta Hamiltonian was
calculated by Fal'ko and Efetov.~\cite{kn:falkoefetov1996}

\subsection{Weak magnetic field}
\label{sec:wmf}
Using Eq.~(\ref{Pc}), the calculation of the peak-height correlation
function $P_c(g_{\mu},g_{\nu})$ becomes a matter of
quadrature. Closed-form results can be obtained for the case $\alpha
\gg 1$:
\begin{eqnarray} \label{correlGOEAsym}
  P_c(g_\mu,g_\nu) =
  \frac{1}{2\alpha^2} {\rm e}^{-g_{\mu} - g_{\nu}}
  (1 - g_{\mu}) (1 - g_{\nu})
\end{eqnarray}
for highly asymmetric tunneling contacts ($T_L \ll T_R$), whereas for symmetric contacts ($T_L = T_R$)
\begin{eqnarray}
  P_c(g_{\mu},g_{\nu}) &=&
  \frac{1 }{16 \alpha^2} 
  W(g_{\mu}) W(g_{\nu}),\label{correlGOESym}
\end{eqnarray}
where the function $W$ is a linear combination involving modified Bessel functions,
\begin{eqnarray}
  W(g) = 2 g \;{\rm e}^{-g} \left[(2-2g) K_0(g) + (1-2g) K_1(g) \right]. \label{eq:W}
\end{eqnarray}
The degree of correlation is well characterized by the first moment of $P_c(g_\mu,g_\nu)$,
\begin{equation}
  C_{\mu\nu} = \langle g_{\mu} g_{\nu} \rangle -
  \langle g_{\mu} \rangle \langle g_{\nu} \rangle.
\end{equation}
In the regime $\alpha\gg 1$ we find from Eqs. (\ref{correlGOEAsym}) and (\ref{correlGOESym})  
\begin{subequations}
\begin{eqnarray}
  C_{\mu \nu} &=& \frac1{2\alpha^2} \ \
  \mbox{if $T_L \ll T_R$}, \\
  C_{\mu \nu} &=& \frac1{9 \alpha^2} \ \
  \mbox{if $T_L = T_R$}.
  \label{Clargealpha}
\end{eqnarray}
\end{subequations}

For very weak magnetic fields, $\alpha \ll 1$, evaluation of
the correlator $P_c$ requires knowledge of the $|\mu-\nu|$-level
spacing distribution functions for the Gaussian Orthogonal Ensemble
of random-matrix theory. Although the solution to this problem is 
known in the form of a product of eigenvalues of a certain integral equation,~\cite{kn:mehta} no closed-form expressions exist to the best of our knowledge. Moreover, if the energy levels $\mu$ and $\nu$ are nearest neighbors, small-$\alpha$ perturbation theory fails for small level separations. (Upon averaging over energy, Eq.~(\ref{rhovar}) gives a logarithmic divergence.)
This problem can be circumvented, noting that the orthogonal invariants $\rho_{\mu\nu}$, the only source of correlations if $\alpha\ll 1$, are completely determined by properties of the two energy levels under consideration. Therefore, the peak-height correlations may
be calculated using a $2 \times 2$ Hamiltonian instead of
the full $N \times N$ random matrix. In the eigenvector basis of the Pandey-Mehta Hamiltonian (\ref{pandey}) at $\alpha=0$, the
appropriate two-level Hamiltonian reads
\begin{eqnarray}
\label{twolevel}\mathcal{H}=\left(\begin{array}{cc} 
\varepsilon_\mu & i\alpha A_{\mu\nu} /\sqrt{2N}\\
i\alpha A_{\nu \mu}/\sqrt{2N}&\varepsilon_\nu
\end{array}\right),
\end{eqnarray}
where  $\varepsilon_{\mu}$ and $\varepsilon_{\nu}$ 
are the two energy levels at $\alpha=0$, and $A_{\mu\nu} = - A_{\nu\mu}$ is the corresponding
matrix element of the perturbing matrix $A$, cf. Eq.~(\ref{pandey}).
Solving for the eigenvectors of ${\cal H}$ and calculating the 
distribution of $\rho_{\mu\nu}$ exactly, we were able
to compute the small-$\alpha$ behavior of the
correlator $C_{\mu\nu}$ for $\nu = \mu+1$,
\begin{subequations}
 \label{GOEFirst}
\begin{eqnarray}
  C_{\mu\nu}&\approx &
  \frac{\alpha^2}{2\pi} \ln \frac{4 \pi}{\alpha^2 {\rm e}^2} \qquad\;\;\,
  \mbox{if $T_L \ll T_R$}, \\
  \label{GOEFirstSym} C_{\mu\nu}& \approx & \frac{3\alpha^2}{16\pi} 
  \ln\frac{0.569}{\alpha^2} \qquad
  \mbox{if $T_L = T_R$}. 
\end{eqnarray}
\end{subequations}
The numerical coefficients inside the logarithms in Eqs.~(\ref{GOEFirst}) were obtained by making use of the Wigner surmise $P(s) = (\pi s/2) \exp(-\pi s^2/4)$ as a numerical approximation to the distribution of nearest-neighbor spacings $s = |\varepsilon_{\mu+1}-\varepsilon_{\mu}|/\Delta$.~\cite{kn:mehta}
\begin{figure}
\scalebox{0.5}{\includegraphics{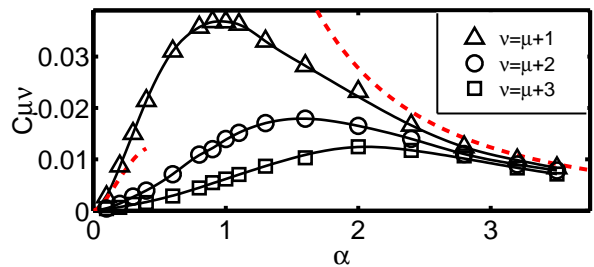}}
\caption{ \label{RMTSimulation} The correlator $C_{\mu\nu}$ for
energy levels $\mu$ and $\nu=\mu+1$, $\mu+2$, $\mu+3$ 
for symmetric tunneling contacts, $T_L=T_R$. The data points are the result of numerical diagonalizations of the
Pandey-Mehta Hamiltonian (\protect\ref{pandey}). The solid curves are drawn as a guide to the eye. The dashed lines
show the large-$\alpha$ and small-$\alpha$ asymptotes (\protect\ref{Clargealpha}) and (\protect\ref{GOEFirstSym}),
respectively.}
\end{figure}

A comparison of our results with the result of numerical diagonalizations of the Pandey-Mehta Hamiltonian is shown in Fig.~\ref{RMTSimulation} for the case of symmetric tunneling contacts. We used random matrices of sizes $N=100$, $200$, and $400$ and extrapolated to $N \to \infty$ to eliminate finite-$N$ effects. Note that throughout the magnetic-field range of interest, correlations between peaks are positive: small peaks are more likely to be surrounded by small peaks, and large peaks attract more large peaks.

\subsection{Different large magnetic fields}
\label{sec:lmf}
\begin{figure}
\scalebox{0.5}{\includegraphics{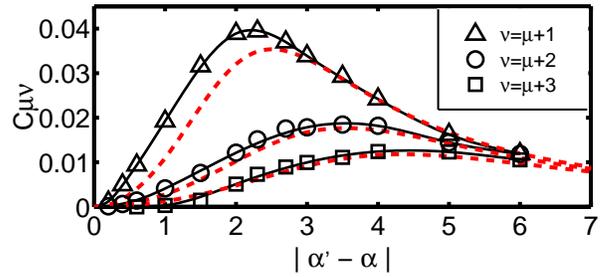}}
\caption{ \label{RMTSimulationGUE} 
The correlator $C_{\mu\nu}$ for different conductance peak
heights at different values of the magnetic field, for
energy levels $\mu$ and $\nu=\mu+1$, $\mu+2$, $\mu+3$ and
symmetric tunneling contacts, $T_L=T_R$. The data points are the result of numerical diagonalizations of the Pandey-Mehta Hamiltonian~(\protect\ref{pandey}). The solid curves are drawn as a guide to the eye. The dashed curves show Eq.~(\protect\ref{eq:cprime}) with $\epsilon_\mu-\epsilon_\nu=(\mu-\nu)\Delta$, which is asymptotically correct for large $|\alpha'-\alpha|$ or large $|\mu-\nu|$.}
\end{figure}

We now turn to correlations between conductance peaks at {\em different large} values of the magnetic field. In particular, we are interested in the connected part of the joint distribution $P(g_{\mu},g'_{\nu})$ where $g_{\mu}$ is a (dimensionless) peak height at magnetic-field strength $\alpha$, while $g_{\nu}$ is a peak height of a different orbital state at a different magnetic-field strength $\alpha'$. (Magnetic-field autocorrelations for the same peak were studied by Bruus {\em et al.} in Ref.~\onlinecite{kn:bruus}) 
Peaks corresponding to different wavefunctions are uncorrelated if measured at the same value of the magnetic field, but correlated at different values of the magnetic field. In order to describe these correlations, we still employ the Pandey-Mehta Hamiltonian~(\ref{pandey}) but now take $\alpha$ and $\alpha'$ large. The joint distribution of eigenvectors $v_{\mu}$ and $v_{\nu}$ at different values of $\alpha$ is thus characterized by the unitary invariants
\begin{equation}
  \tilde \rho_{\mu\nu} = v_{\mu}^{\dagger}(\alpha) v_{\nu}(\alpha'),
\end{equation}
where $v_{\mu}(\alpha)$ denotes the eigenvector of the $\mu$th
level at magnetic-field strength $\alpha$.
At fixed $\tilde{\rho}_{\mu\nu}$, the eigenvector 
components are distributed according to a multivariate Gaussian 
distribution with covariance matrix determined by the 
pair correlator,
\begin{eqnarray}
  \label{variancesGUE}
  \langle v_{\mu,m}^*(\alpha) v_{\nu,n}(\alpha')\rangle_{\tilde{\rho}} 
  = \frac{\delta_{mn}\tilde{\rho}_{\mu\nu}}{N}.
\end{eqnarray}
The second moments $\langle|\tilde{\rho}_{\mu\nu}|^2\rangle$ are known 
in the regimes $|\varepsilon_\mu-\varepsilon_\nu| \gg
\min(\Delta,|\alpha'-\alpha|\Delta)$ or $|\alpha'-\alpha|\gg 1$,~\cite{kn:wilkwalk}
\begin{eqnarray}
\label{rhovartilde} \langle |\tilde{\rho}_{\mu\nu}|^2\rangle &=&
\frac{2(\alpha'-\alpha)^2}{(\alpha'-\alpha)^4+4\pi^2(\varepsilon_\mu-\varepsilon_\nu)^2/\Delta^2}.
\end{eqnarray}
The remainder of the calculation proceeds as before, the only difference being the slightly different expression for the average $\langle |\tilde{\rho}_{\mu\nu}|^2\rangle$ in this case. We thus obtain
\begin{eqnarray}
  P_c(g_{\mu},g_{\nu}') =
  \frac{2(\alpha'-\alpha)^2 (1 - g_{\mu})(1-g_{\nu})}
  {(\alpha'-\alpha)^4+4\pi^2(\varepsilon_\mu-\varepsilon_\nu)^2/\Delta^2}
  {\rm e}^{-g_{\mu} - g_{\nu}'}\;\;&&
\end{eqnarray}
in the limit $T_L \ll T_R$, whereas
\begin{eqnarray}
  P_c(g_{\mu},g_{\nu}') = \frac14 
  \frac{(\alpha'-\alpha)^2 
  W(g_{\mu}) W(g_{\nu}')}
  {(\alpha'-\alpha)^4+4\pi^2(\varepsilon_\mu-\varepsilon_\nu)^2/\Delta^2} 
\end{eqnarray} 
for symmetric tunneling contacts, with the function $W(g)$ as defined in Eq.~(\ref{eq:W}). For the correlator
$C_{\mu\nu} = \langle g_{\mu} g'_{\nu}\rangle
- \langle g_{\mu} \rangle \langle g'_{\nu} \rangle$, 
this implies
\begin{subequations}
\begin{eqnarray}
 C_{\mu\nu}
  = \frac{2(\alpha'-\alpha)^2}
  {(\alpha'-\alpha)^4+4\pi^2(\varepsilon_\mu-\varepsilon_\nu)^2/\Delta^2}&&\mbox{if $T_L \ll T_R$},\nonumber\\\\
 C_{\mu\nu}  = \frac{4(\alpha'-\alpha)^2}
  {9 (\alpha'-\alpha)^4+36\pi^2(\varepsilon_\mu-\varepsilon_\nu)^2/\Delta^2}&&\mbox{if $T_L=T_R$}.  \nonumber \\\label{eq:cprime}
\end{eqnarray}
\end{subequations}
For $|\alpha'-\alpha| \gg 1$ this
result is also valid for the case $\mu=\nu$ and agrees
with previous work by Bruus {\em et al.} in Ref.~\onlinecite{kn:bruus}. 
In Fig.~\ref{RMTSimulationGUE}, we compare $C_{\mu\nu}$
to numerical diagonalizations of the Pandey-Mehta Hamiltonian (\ref{pandey}), using random matrices of sizes $N=100$, $150$, and $300$ with
extrapolation to $N \to \infty$ to eliminate finite-$N$ effects.\\

\section{Application to Spin-Orbit Scattering and Spectral Scrambling}
\label{sec:appl}
Although our calculations were performed for peak height correlations that resulted from an external magnetic field or a change in the external magnetic field, they can also be of relevance as an effective description of correlations due to spin-orbit scattering in GaAs quantum dots or due to ``spectral scrambling''.

In two-dimensional GaAs quantum dots, spin-orbit scattering is described by the following two-dimensional effective Hamiltonian:~\cite{kn:alfalko}
\begin{eqnarray}
H_\mathrm{SO}=\frac1{2m}\left[\frac{p_2\sigma_1}{\lambda_2}-\frac{p_1\sigma_2}{\lambda_1}\right].\nonumber
\end{eqnarray}
Here, $\sigma_i$ are Pauli-matrices, and $\lambda_1$, $\lambda_2$ are length scales associated with spin-orbit coupling along the directions $\hat{x}_1$ and $\hat{x}_2$ that span the plane in which the dot is formed. In the limit where $\lambda_{1}$, $\lambda_2$ are large compared to the linear dot size $L$, the spin-orbit contribution can be mapped onto an effective magnetic field $\vec{B}_{\rm SO}$ by means of a suitable unitary tranformation of $H_\mathrm{SO}$:~\cite{kn:alfalko}
\begin{eqnarray}
\tilde{H}_\mathrm{SO}=\frac1{2m}\left(\vec{p}-\vec{\mathrm{a}}_\perp\right)^2 -\frac{\vec{p}^2}{2m},
\end{eqnarray}
where 
\begin{eqnarray}
\vec{\mathrm{a}}_\perp=\frac{\sigma_3}{4\lambda_1\lambda_2} [\hat{x}_3\times\vec{r}].
\end{eqnarray}
is the vector potential that generates the leading spin-orbit effect. Hence, weak spin-orbit scattering takes the form of an effective magnetic field $B_{\rm SO}=\hbar c/2e\lambda_1\lambda_2$ of opposite sign for the two spin directions and perpendicular to the plane of the two-dimensional electron gas in which the dot is formed. The parameter $\alpha$ in the Pandey-Mehta Hamiltonian (\ref{pandey}) is then given by
\begin{eqnarray}
\alpha = \pm\gamma \frac{V}{4\pi\lambda_1\lambda_2} \sqrt{\frac{E_T}{ \Delta}},
\end{eqnarray}
where the $\pm$ corresponds to the two spin directions, and $\gamma$ is the same geometric factor as in Section \ref{sec:model}. Experimental estimates suggest $\alpha \lesssim 1$, which implies that the effective magnetic field is weak enough to only partially break time-reversal symmetry.~\cite{kn:zumbuhl} 

The peak-height correlations for a weak magnetic field calculated in Section \ref{sec:wmf} thus provide a good description of intrinsic peak-height correlations for a quantum dot with weak spin-orbit scattering in the absence of an external magnetic field. On the other hand, when a large external magnetic field $B$ is applied perpendicular to the dot, electrons move in different effective magnetic fields $B \pm B_{\rm SO}$, depending on the direction of their spin. At zero temperature, conductance peaks correspond to resonant tunneling for one of the two spin directions. Our calculations in Section \ref{sec:lmf} show that peaks originating from resonances with the same spin direction, i.e. $|\alpha'-\alpha|=0$, will have uncorrelated heights, whereas peaks originating from resonances with opposite spin will have correlated heights, corresponding to the case of large magnetic fields with magnetic-field strength difference $|\alpha'-\alpha|=\gamma ({V}/{2\pi\lambda_1\lambda_2}) \sqrt{{E_T}/{ \Delta}}$.

``Scrambling'' is the effect that each electron added to the
quantum dot causes a small change to the self-consistent
potential in the dot.~\cite{kn:alhassid99,kn:scrambling} 
Hence, every conductance peak is taken at a slightly different 
realization of the dot's potential. While this
leads to a {\em decorrelation} of peak heights corresponding to
the same orbital state, scrambling also causes a positive
correlation between peak heights corresponding to different
orbital states, as we have shown above for the case of a large
applied magnetic field. (In the unitary ensemble, a change in potential has the same effect as a change in the applied magnetic field. The situation at zero applied magnetic field would correspond to the orthogonal ensemble of random-matrix theory, for which the calculation proceeds along the same lines and gives
similar results.~\cite{kn:braig}) The effect of adding $n$ electrons to a disordered quantum dot corresponds to a parameter change $|\alpha'-\alpha| \sim n
\sqrt{\Delta/E_{\rm T}}$,~\cite{kn:scrambling} 
where $E_{\rm T}$ is the Thouless 
energy.
Hence, we conclude from our calculations that the resulting correlations between peak heights are of
order $n^2 \Delta/E_{\rm T}$. While such correlations may be of 
numerical importance, its dependence on the ratio $E_{\rm T}/\Delta$ is 
the same as that of the
non-universal peak-height correlations in a disordered
dot.~\cite{kn:mirlin} We therefore see that both types of correlations need to be taken into account for a complete understanding of spectral scrambling effects.

\begin{acknowledgments}
We thank Harold Baranger, Henrik Bruus,
Josh Folk, Mikhail Polianski, and Gonzalo Usaj for discussions.
This work was supported by the Cornell Center for
Materials Research under NSF grant no.\ DMR0079992, 
by the NSF under grant no.\ DMR
0086509, and by the Packard foundation.
\end{acknowledgments}


\end{document}